# High performance on-demand de-identification of a petabyte-scale medical imaging data lake

Joseph Mesterhazy, Garrick Olson, Somalee Datta

Research IT, Technology & Digital Solutions, Stanford Health Care

## Abstract:

With the increase in Artificial Intelligence driven approaches, researchers are requesting unprecedented volumes of medical imaging data which far exceed the capacity of traditional on-premise client-server approaches for making the data research analysis-ready. We are making available a flexible solution for on-demand de-identification that combines the use of mature software technologies with modern cloud-based distributed computing techniques to enable faster turnaround in medical imaging research. The solution is part of a broader platform that supports a secure high performance clinical data science platform.

## Background:

In order to support secondary use of data generated during the course of clinical care to advance the science and practice of medicine, Stanford Medicine is developing its next generation clinical data platform for research and related data science ecosystem [Datta2020]. This platform is an ecosystem consisting of a data lake (STanford medicine Research data Repository or STARR), a suite of analytical data warehouses and cohort building applications and a secure data science facility. A sampling of medical imaging research shows that researchers have been successful at Stanford [Varma2019, Patel2019, Irvin2019, Park2019, Dunnmon2019, Bien2018, Banerjee2018a, Banerjee2018b, Rajpurkar2018, Esteva2017] and elsewhere [Peng2020, Zhang2020, Gulshan2016], in applying AI approaches to imaging data to derive scientific insights. In parallel, national experts are charting the path of AI in medical imaging [Allen2019]. We expect AI research in imaging to become ubiquitous resulting in even greater demand for imaging data at Academic Medical Centers like ours. As a result, our new ecosystem has a strong focus on supporting our imaging community.

Modern hospital environments are generating hundreds of terabytes of new imaging data per year. At Stanford, the radiology imaging data is growing at the rate of 450 terabytes annually representing various different modalities (Figure 1) and often dozens of manufactures and hundreds of devices. The primary challenges to delivering medical imaging data for research purposes are acquisition, de-identification and delivery. The DICOM medical imaging standard [Dicom2019] is widely supported by the open-source community with multiple software packages available for de-identification [Aryanto2015]. The MIRC Clinical Trial Processor (MIRC CTP) [RSNA2019] is a popular and performant [Aryanto2015] open-source software package for conducting clinical trials with DICOM image data, and includes a comprehensive DICOM anonymizer plugin.

Traditional approaches for de-identifying radiology medical images involve the use of one or more software applications directly connected to a clinical imaging archive. When imaging studies are identified for research use, the anonymization software retrieves and de-identifies the images from the clinical archive. This approach quickly breaks down in scale when multiple researchers are requesting billions of images and hundreds of terabytes of data.

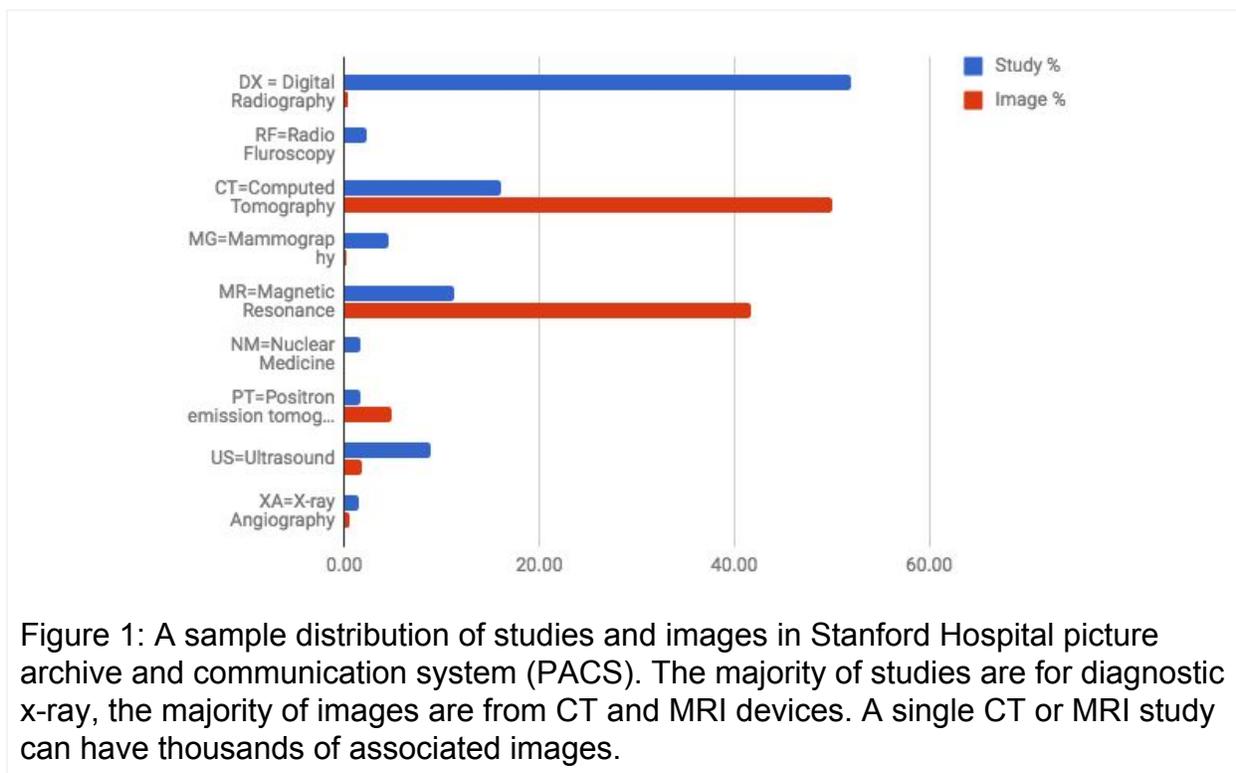

Figure 1: A sample distribution of studies and images in Stanford Hospital picture archive and communication system (PACS). The majority of studies are for diagnostic x-ray, the majority of images are from CT and MRI devices. A single CT or MRI study can have thousands of associated images.

Recently cloud computing providers have begun offering API services for text and medical image de-identification, however they were incompatible with our requirement for reproducibility and control over the de-identification process. Vendor API's operate as a "black box" that changes without notice, affecting reproducibility. These APIs are in early stages, and can lack functionality like complete control over the replacement of DICOM attributes. These APIs are reassessed periodically as they mature and provide greater flexibility, however their proprietary nature, lack of change control and vendor lock-in will always present a major challenge for implementation.

Another challenge to delivering medical image data for research purposes is that Stanford University Privacy Office (UPO) determines that algorithmically de-identified DICOM images have the potential for leaking incidental Protected Health Information (PHI). In order to confirm that the de-identified data has no incidental PHI, UPO requires the algorithmic output to be scanned by humans. While such expert determination is used for broader data sharing (e.g., https://aimi.stanford.edu/research/public-datasets), it is not feasible to scan each image by eye for internal research use with hundreds of terabytes of data. To address security and patient privacy at Stanford, we have built a secure data science facility [Datta2020] to support downstream data processing on petascale data. To support research needs including the use of modern deep learning approaches, we present a new de-identification system that combines the reliability of existing de-identification software with the scale of cloud-based distributed computing. We also decouple the data processing system from the clinical imaging archive to reduce any adverse impact on clinical operations. The new imaging pipeline acquires data on-demand from the clinical radiology archive, algorithmically de-identifies the data in parallel and delivers to the researcher's secure data science workspace.

## Method:

When de-identifying patient data, patient identifiers must be replaced by unique, anonymized codes, a process known as pseudonymization [Noumeir2007]. Our de-identification pipelines support two separate research stages. First one is pre-IRB non-human subject research. The de-identification aggressively strips everything that may have HIPAA identifiers. In this mode, the anonymized codes generated for non-human subject research can never be reversed and linked to identified patient data. Second is for post-IRB study, the researcher can request de-identified images which meet the HIPAA minimum necessary for the study and can additionally request links between the anonymized images and the original patient identifiers.

We have developed an on-demand highly scalable de-identification solution in lieu of a single archive of permanently de-identified images to support the two distinct use cases. The on-demand pipeline also allows us to make continuous improvements to de-identification quality (i.e., reduce PHI leak) unlike a single permanently de-identified image archive. We will show in the Results section that our cost of de-identification is extremely low to support this on-demand workflow.

Specifically, we have written a distributed software application (STARR-Radiology aka STARR-Radio) which a) forwards fully-identified DICOM image data from on-premise clinical systems to STanford medicine Research data Repository (aka STARR data lake), specifically, to an encrypted and distributed cloud object storage service, b) listens for de-identification requests using a publish/subscribe messaging model, c) instantiates an appropriate number of de-identification compute instances based on the size of the message queue, and d) delivers the de-identified DICOM images to an object store accessible to the researcher for downstream data science. Although CTP is designed for installation as a client-server application, it's highly extensible nature allowed us to extract only the DICOM filtering and anonymizing components for our use. Additionally, MIRC CTP uses the dcm4che [Dcm4che2019] libraries, which provide support for acquiring, manipulating, and transmitting DICOM files. These two software components were integrated with the cloud platform that underlies the STARR data lake.

A central database and server component was created to store workflow information relevant to the lifetime of a de-identification request. When a request for images has IRB approval, an entry for the approved study is inserted in the database along with associated accession numbers (representing specific imaging studies for that study). The accession numbers are first validated as eligible for research, and if validated a new anonymized accession number, patient MRN, and randomized date jitter specific to the specific research study are created. These identifiers, combined with anonymization rules appropriate for the associated IRB, are transmitted to a central messaging queue. An auto-scaling compute pool which is subscribed to the messaging queue creates an appropriate number of compute instances based on the total number of outstanding messages in the queue and the expected delivery window. Each worker retrieves messages from the queue, downloads and de-identifies the DICOM files using the MIRC anonymization engine, and uploads the de-identified images to an object store accessible to the researcher. Compute instances are deleted once the message queue is empty, and a manifest file is created which indicates the transformations applied to each image, along with success or failure states.

The greatest challenge encountered was creating and validating rules for reliably de-identifying DICOM imaging studies. While the MIRC anonymizer includes a set of rules intended as a starting point for further development, they are not acceptable in their default state [Aryanto2015]. The DICOM anonymization process (Figure 2a) is executed in three discrete stages: filtering, scrubbing, and anonymization. The filtering stage accepts or discards the image based on metadata rules. For example, the filtering stage removed encapsulated PDF reports, scanned intake forms, outside study alerts etc. If the image is not discarded, it is then passed to the scrubbing stage, which removes rectangular regions from the image which are known to contain PHI. Here, the regions of the image containing PHI are replaced by black pixels and recompressed using the JPEG Lossless syntax. The rules used for pixel blanking can be found on GitHub (https://github.com/susom/mirc-ctp/blob/master/stanford-scrubber.script). The final stage removes or replaces image metadata which is known to contain PHI.

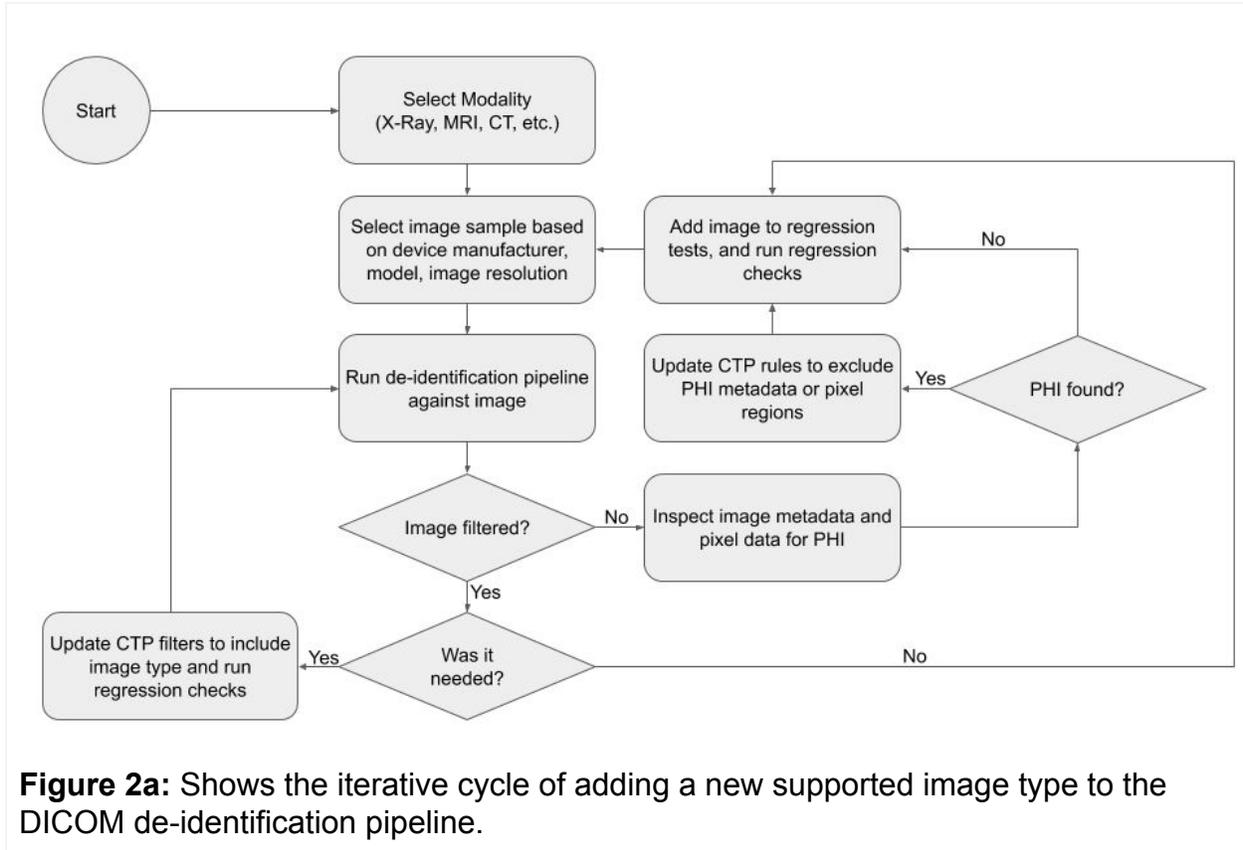

**Figure 2a:** Shows the iterative cycle of adding a new supported image type to the DICOM de-identification pipeline.

Using the DICOM Basic Application Confidentiality Profile with the Clean Graphics and Retain Longitudinal Temporal Information With Modified Dates options as a starting point (http://dicom.nema.org/dicom/2013/output/chtml/part15/sect_E.2.html), new filtering, anonymization and pixel scrubbing rules were created using an iterative process. Images for each modality are categorized by make, model and image resolution. A sample for each combination is executed against the pipeline, which is modified as necessary to remove PHI. After every iteration the image is added to the regression suite, which is then executed to ensure new rules do not break existing ones.

During development regression tests were created as new rules were added to ensure new rules did not break existing rules. The regression tests are written using the Cucumber testing framework (https://cucumber.io/) and are human readable (Figure 2b).

```
Feature: Anonymize CT images filtering where appropriate

 Background:
  Given the pipeline uses the anonymizer
script,"stanford-anonymizer.script"
  Given the pipeline uses the pixel script, "stanford-scrubber.script"
```

```
    Given the pipeline uses the filter script, "stanford-filter.script"
    And script parameter "accession" is "ACN123"
    And script parameter "mrn" is "MRN123"
    And script parameter "jitter" is "-6"

Scenario: All files in the PT/Anonymize folder should be anonymized
    Given the DICOM directory "dicom-phi/PT/Anonymize"
    When ran through the deid pipeline
    Then the images SHOULD be anonymized

Scenario: REG-PCT01 GE PET/CT fusion
    Given the DICOM directory "dicom-phi/PT/Scrub/GE/Discovery/512x512"
    When ran through the deid pipeline
    Then the resulting images should be scrubbed at 256,0,256,22
    And the resulting images should be scrubbed at 300,22,212,80
    And the resulting images should be scrubbed at 10,478,100,10

Scenario: All files in the PT/Filter folder should be filtered
    Given the DICOM directory "dicom-phi/PT/Filter"
    When ran through the deid pipeline
    Then the images SHOULD NOT pass the filter
```

**Figure 2b:** An excerpt of a regression test for PET/CT devices. The first scenario checks that all DICOM files in the directory Anonymize are modified by the anonymizer and jittered correctly. The second scenario inspects rectangular regions of the scrubbed output to ensure they are blank. The third scenario requires all files in the Filter directory to be excluded by the pipeline. If any of these tests fail, the regression test results in failure.

The maturity of the CTP software itself presented unique challenges. While CTP is written in the cross-platform Java language, it requires libraries which utilize the Java Native Interface, and only run on x86 or SPARC platforms. These libraries are no longer supported by Oracle, and posed problems during development, requiring the use of Docker containerization as a compatibility layer. Additionally, the CTP software uses older Java technologies, such as file references instead of stream buffers. This requires the de-identification workers to store images in block storage attached to the OS, instead of networked direct data streams. The negative performance impact of block storage was mitigated by mounting system RAM as block storage (aka ramdisk) which is used as ephemeral storage during de-identification.

## Results:

In this section, we present the performance achieved for the de-identification pipeline. A selection of random studies (containing one or more images) for three different imaging modalities were passed through the de-identification pipeline consisting of 8 compute instances running Google Container Optimized OS, each with 32 CPUs and 208GB of RAM. The de-identification application is executed within docker, using a 'tmpfs' mount as a high-performance ramdisk. Anonymized instances which have not been filtered are uploaded to Google Cloud Storage and deleted from ramdisk.

| Studies / Modality | Filtered Instances | Anonymized Instances | Scrubbed Instances | Total Bytes | Duration | Aggregate Throughput | Cost |
|---|---|---|---|---|---|---|---|
| 5,000 CT | 2,663,139 | 4,997,888 | 56,318 | 3 TB | 45 min | 1.25 GB/s | $5.68 |
| 10,000 US | 70,550 | 7,390 | 420,734 | 3.5 TB | 60 min | 977 MB/s | $8.52 |
| 100,000 X-Ray | 151,096 | 212,882 | 3,662 | 2.3 TB | 56 min | 684 MB/s | $7.95 |

**Table 1:** Performance benchmarks for the de-identification of a random selection of imaging studies.

The low cost of de-identification allows us to store a single copy of the identified image and provide on-demand de-identification as needed.

## Discussion:

During the iterative development process it was found that certain types of images needed to be excluded completely as the probability for PHI leakage was high. Such examples include:

1. Analog x-ray film which has been digitized by a film scanner. For example, all images with manufacturer "Vidar" are filtered by the pipeline. Scanned analog x-rays are highly problematic as PHI may be present at any location on the film (in the form of a stick or hand-written notes) and the image may be the incorrect orientation, eg. flipped or upside-down.
2. DICOM files containing:
    a. Encapsulated PDF documents
    b. Structured Report (SR) documents
    c. Presentation state objects
    d. Uncommon modality attributes (eg. RAW)
    e. Secondary capture objects*
    f. Burned-in annotation is set to "Yes"*
    g. ConversionType attribute is an empty-string
    h. ImageType attribute contains DERIVED or SECONDARY*
3. DICOM from video-capture devices

* May be bypassed by specific whitelisting rules based on other attributes.

Ultrasound devices in particular were found to be the most complex for reliable de-identification, followed by digital x-ray. While other modalities may occasionally contain PHI in the pixel data (for example CT radiation dose exposure screens), ultrasound almost always contains PHI in the pixel data. It is common for multiple regions within a single image to have pixel data that needed to be removed, which may be different even within the same make and model of an ultrasound device depending on the resolution of the image. For this reason ultrasound devices are only supported via a whitelist: if there is no rule for a specific make, model, and image resolution for an ultrasound image, the image is filtered.

| Ultrasound make | Models | Resolution variations |
|---|---|---|
| GE | 35 | 151 |
| Siemens | 13 | 24 |
| Acuson | 2 | 14 |
| Philips | 12 | 22 |
| Toshiba | 13 | 24 |
| SonoSite | 6 | 7 |
| Zonare | 3 | 4 |
| BK Medical | 3 | 7 |
| Aloka | 7 | 10 |
| SuperSonic Imaging | 1 | 15 |
| Samsung | 8 | 16 |

**Table 2:** Ultrasound manufacturers and the number of image variations per manufacturer found in Stanford data. An image variation represents the total number of combinations of model and image resolutions. For example 38 different image resolutions were found for the GE (make) LOGIQE9 (model) ultrasound device alone. Each resolution required a new image scrubbing rule, for a total of 151 different rules for GE devices alone. This table represents 99% of the manufacturers in the clinical imaging archive.

It should be noted that certain data types, such as head images, even when de-identified fully (i.e., no incidental PHI) could be combined with other publicly available data to re-identify the original patient [Schwarz2019]. There are advanced techniques for skull stripping [Kalavathi2016] that should be applied if research requires public sharing of head imaging data.

## Future work:

Future work focuses on several different areas. Firstly, we are building a DICOM metadata store using Google BigQuery [Datta2020] that will host metadata from raw identified DICOM images. A pre-IRB de-identified version of this store will be made accessible for cohort development. Secondly, we are leveraging our clinical text mining framework [Datta2020] to augment cohort development e.g. text mining radiology reports. Thirdly, we hope to integrate OCR and other machine learning approaches to improve image de-identification. Finally, we plan to generate a pre-IRB non-human subject dataset containing a stratified subset of imaging data that is linkable to the pre-IRB OMOP clinical data [Datta2020] for research acceleration.

# Acknowledgments:

STARR suite is made possible by Stanford School of Medicine Dean's Office.

Using CRediT taxonomy (https://casrai.org/credit/), we present the contributing roles for our authors - Joseph Mesterhazy (Conceptualization, Methodology, Data curation, Formal Analysis, Investigation, Software, Validation), Garrick Olson (Supervision, Resources, Investigation) and Somalee Datta (Writing - review & editing). For the broader team, we are grateful to Nivedita Shenoy (Project administration), Agile Program Manager, Technology & Digital Solutions, who helped our engineering team with implementation of the scrum process during the development cycle. We are grateful to the feedback from early adopters at the Center for Artificial Intelligence in Medicine and Imaging (https://aimi.stanford.edu/). We acknowledge support of members of Research IT team (https://med.stanford.edu/researchit), the broader Technology & Digital Solutions (https://tds.stanfordmedicine.org/) IT team, and Michael Halaas (Funding acquisition), Deputy Chief Information Officer of Technology & Digital Solutions and Associate Dean of Industry Relations at Stanford School of Medicine.

10 of 116. [Datta2020] https://arxiv.org/abs/2003.10534
7. [Dcm4che2019] Dcm4che, a collection of open source applications and utilities for the healthcare enterprise. Available via https://www.dcm4che.org/. Accessed December 2019.
8. [Dicom209] N. E. M. A. (NEMA), "The DICOM Standard." [Online]. Available via http://medical.nema.org/. Accessed December 2019.
9. [Dunnmon2019] Dunnmon JA, Yi D, Langlotz CP, Ré C, Rubin DL, Lungren MP. Assessment of Convolutional Neural Networks for Automated Classification of Chest Radiographs. Radiology. 2019 Feb;290(2):537-544.
10. [Esteva2017] Esteva A, Kuprel B, Novoa RA, Ko J, Swetter SM, Blau HM, Thrun S, Dermatologist-level classification of skin cancer with deep neural networks, Nature. 2017 Feb 2;542(7639):115-118. doi: 10.1038/nature21056. Epub 2017 Jan 25.
11. [Gulshan2016] Gulshan V, Peng L, Coram M, Stumpe MC, Wu D, Narayanaswamy A, Venugopalan S, Widner K, Madams T, Cuadros J, Kim R, Raman R, Nelson PC, Mega JL, Webster DR, Development and Validation of a Deep Learning Algorithm for Detection of Diabetic Retinopathy in Retinal Fundus Photographs, JAMA. 2016 Dec 13;316(22):2402-2410. doi: 10.1001/jama.2016.17216.
12. [Irvin2019] Irvin J, Rajpurkar P, Ko M, Yu Y, Ciurea-Ilcus S, Chute C, Marklund H, Haghgoo B, Ball R, Shpanskaya K, Seekins J, Mong DA, Halabi SS, Sandberg JK, Jones R, Larson, DB, Langlotz CP, Patel BN, Lungren MP, Ng AY. CheXpert: A Large Chest Radiograph Dataset with Uncertainty Labels and Expert Comparison. AAAI. 2019;590-597.
13. [Kalavathi2016] Kalavathi P, Prasath VBS, Methods on Skull Stripping of MRI Head Scan Images—a Review, J Digit Imaging. 2016 Jun; 29(3): 365–379.
14. [Noumeir2007] Noumeir R, Lemay A, Lina JM. Pseudonymization of radiology data for research purposes. J Digit Imaging. 2007;20(3):284–295. doi:10.1007/s10278-006-1051-4
15. [Park2019] Park A, Chute C, Rajpurkar P, Lou J, Ball RL, Shpanskaya K, Jabarkheel R, Kim LH, McKenna E, Tseng J, Ni J, Wishah F, Wittber F, Hong DS, Wilson TJ, Halabi S, Basu S, Patel BN, Lungren MP, Ng AY, Yeom KW. Deep Learning-Assisted Diagnosis of Cerebral Aneurysms Using the HeadXNet Model. JAMA Netw Open. 2019 Jun 5;2(6):e195600.
16. [Peng2020] Peng Z, Fang X, Yan P, Shan H, Liu T, Pei X, Wang G, Liu B, Kalra MK, Xu XG, A Method of Rapid Quantification of Patient-Specific Organ Doses for CT Using Deep-Learning based Multi-Organ Segmentation and GPU-accelerated Monte Carlo Dose Computing, Med Phys. 2020 Mar 10. doi: 10.1002/mp.14131. [Epub ahead of print]
17. [Patel2019] Patel BN, Rosenberg L, Willcox G, Baltaxe D, Lyons M, Irvin J, Rajpurkar P, Amrhein T, Gupta R, Halabi S, Langlotz C, Lo E, Mammarappallil J, Mariano AJ, Riley G, Seekins J, Shen L, Zucker E, Lungren MP, Human–machine partnership with artificial intelligence for chest radiograph diagnosis. npj Digit. Med. 2, 111 (2019) doi:10.1038/s41746-019-0189-7